\documentclass[materials,article,submit,pdftex,moreauthors]{Definitions/mdpi}
\firstpage{1} 
\pubvolume{1}
\issuenum{1}
\articlenumber{0}
\pubyear{2022}
\copyrightyear{2022}
\datereceived{} 
\dateaccepted{} 
\datepublished{} 
\hreflink{https://doi.org/} 
\pdfoutput=1

\Title{Energy relaxation and electron-phonon coupling in laser-excited metals}

\Author{Jia Zhang$^{1,2}$, Rui Qin$^{2}$, Wenjun Zhu$^{2}$, Jan Vorberger$^{1}$*\orcidA{}}

\address{%
$^{1}$ \quad Institute of Radiation Physics, Helmholtz-Zentrum Dresden-Rosendorf, Bautzner Landstraße 400, 01328 Dresden, Germany\\
$^{2}$ \quad National Key Laboratory of Shock Wave and Detonation Physics, Institute of Fluid Physics, China Academy of Engineering Physics,
Mianyang 621999, China}

\corres{Correspondence: j.vorberger@hzdr.de}

\abstract{The rate of energy transfer between electrons and phonons is investigated by a first principles framework for electron temperatures up to $T_e=50000$~K while considering the lattice at ground state. Two typical but differently complex metals are investigated, namely Aluminium and Copper. In order to reasonably take the electronic excitation effect into account, we adopt finite temperature density functional theory and linear response to determine the electron-temperature-dependent Eliashberg function and electron density of states. Of the three branch-dependent electron-phonon coupling strengths, the longitudinal acoustic mode plays a dominant role in the electron-phonon coupling for Aluminium for all temperatures considered here, but for Copper it only dominates above an electron temperature of $T_e=40000$~K. The second moment of the Eliashberg function and the electron phonon coupling constant at room temperature $T_e=315$~K show good agreement with other results. For increasing electron temperatures, we show the limits of the $T=0$ approximation for the Eliashberg function. Our present work provides a rich perspective on the phonon dynamics and this will help to improve insight into the underlying mechanism of energy flow in ultra-fast laser-metal interaction.}

\begin{document}
\nolinenumbers
\section{Introduction}
With the advent of femtosecond pump-probe setups, remarkable progress has been made in the study of ultrashort laser-matter interaction during recent decades~\cite{Macchi:2013,Gamaly:2013,waldecker2015compact,bostedt2016linac,seddon2017short,Hofherr:2020}. Also, huge attention has been paid to understand the fundamental material response dynamics and energy relaxation processes in extreme states far from equilibrium induced by laser irradiation~\cite{ng2012outstanding,Vorberger:2010,dorchies2016non,rethfeld2017modelling}. However, precisely because of the non-equilibrium condition, it is a great challenge to theoretically understand the processes.

Regarding the dynamical response of laser-excited materials, the electrons are accelerated by the laser pulse and thermalize within femtoseconds at a level of several tens of thousands Kelvin while leaving the ions in their initial state. If the initial state is a solid, the lattice is heated due to electron-phonon energy transfer to a new thermal equilibrium over several (tens to hundreds of) picoseconds. During this evolution, many interesting effects such as the ultrafast electron and nonequilibrium phonon dynamics~\cite{rethfeld2002ultrafast,silaeva2018ultrafast,weber2019phonon,obergfell2020tracking,lee2021investigation,chen2021electron, grolleau2021femtosecond,ono2017nonequilibrium,klett2018relaxation}, changes in lattice stability~\cite{recoules2006effect,yan2016different,zhang2018lattice,ben2021structural,ono2021lattice}, phase transitions~\cite{harb2008electronically,giret2014nonthermal,lian2016ab,mo2018heterogeneous,medvedev2020nonthermal,jourdain2021ultrafast} and non-equilibrium electron-phonon interactions~\cite{allen1987theory,vorberger2012theory,waldecker2016electron,sadasivam2017theory,maldonado2017theory,ono2018thermalization,ritzmann2020theory,miao2021nonequilibrium,zahn2021lattice,zahn2021intrinsic} take place. It should be pointed out that these various physical processes are all driven by electronic excitations. For instance, there are two melting mechanisms: the one is traditional thermal melting due to the electron-phonon energy relaxation and the other is non-thermal melting~\cite{harb2008electronically,medvedev2020nonthermal}, which can be attributed to the electronically triggered destabilisation of the lattice under high excitation. In the present work, we concentrate on the microscopic energy flow related to the electron-phonon interaction.

In the case of metals, the phenomenological two temperature model (TTM) proposed by Anisimov {\em et al.} is widely adopted to study the energy relaxation under the excited nonequilibrium conditions~\cite{anisimov1967effect,anisimov1974electron}. In this model, the energy flow evolution is controlled by the electron-phonon coupling factor. Allen {\em et al.} established a microscopic foundation for the TTM and also provided a microscopic expression for the electron-phonon energy exchange rate~\cite{allen1987theory}. Our recent work indicates the limitation of the TTM to predict the lattice dynamics even in a simple metal like Aluminium due to a unique distribution of the branches of the total electron-phonon coupling factor~\cite{waldecker2016electron}. Instead, it becomes necessary in aluminium to separately account for the partial electron-phonon coupling to investigate the energy flow evolution. However, it seems the findings in aluminium cannot be generalized easily as further femtosecond electron diffraction experiments show different results~\cite{zahn2021lattice,zahn2021intrinsic,zahn2021ultrafast}. Thus, accurate electron-phonon coupling factors for a wide variety of elements and materials are in high demand.

In fact, a large number of  theoretical calculations for this important physical quantity have been implemented under different methods for various metals~\cite{vorberger2012theory,waldecker2016electron,zahn2021lattice,zahn2021intrinsic,lin2008electron,mueller2013relaxation,petrov2013thermal,gorbunov2015electron,brown2016ab,medvedev2020electron,medvedev2021contribution,ji2016ab,migdal2016heat,smirnov2020copper,ogitsu2018ab,caro2015adequacy,migdal2015equations,li2022ab,wingert2020direct,milov2018modeling,petrov2020ruthenium}. Nonetheless, from very low electron temperatures($\sim300$~K) to very high electron temperature ($>2000$~K), different models provide estimations varying by a factor of two at the least and up to an order of magnitude~\cite{medvedev2020electron}. 

Notably among these approaches, density functional theory (DFT) is a preferred method~\cite{hohenberg1964inhomogeneous}. This is because the lattice symmetries, ion-ion and electron-ion interaction as well as electron-electron correlations can be taken into consideration naturally. However, standard DFT is only able to compute the electron-phonon coupling within linear response and for systems for which subsystem temperatures can be established. Still, average excitation effects can be accounted for both for the electron as well as for the ion subsystems.
In this paper, we adopt the finite temperature DFT scheme to include the electronic excitation effect into the determination of the electron density of states (DOS), phonon density of states, and Eliashberg function to compute the electron-phonon coupling with all input quantities being fully electron temperature dependent~\cite{hohenberg1964inhomogeneous,mermin1965thermal}.

In the next section \ref{md}, we give some basic theory of the electron-phonon coupling strength and then provide the details of our computations. Section \ref{rd} analyses our predictions and compare them with the existing various theoretical results. Finally we summarise our results and give an outlook on the extension of our approach and also discuss the direct applications using our ab initio determined parameters.
\section{Method}\label{md}
\subsection{Formalism}

Allen used a set of Bloch-Boltzmann-Peierls equations to theoretically investigate the time evolution of a femtosecond laser-excited system of electrons and phonons~\cite{allen1987theory}. It contains the electron-phonon coupling and conserves the total kinetic energy of the electron and phonon subsystems. According to the microscopic essence of the TTM, the electron-phonon energy exchange rate $Z_{ep}$ can be determined by a moment of the Bose distribution function $n_{B}(\omega_{q},t)$ for the phonons~\cite{allen1987theory}
\begin{equation}
\begin{split}
 Z_{ep}(T_{e},T_{l},t)&=\frac{\partial{E_{ph}(t)}}{\partial{t}}
 =\sum_{q}\hbar\omega_{q}(T_{e},T_{l},t)\frac{\partial n_{B}(\omega_{q},t)}{\partial{t}}\\
 &=\frac{4\pi}{N_{c}\hbar}\sum_{q k}\hbar\omega_{q}(T_{e},T_{l},t)|M_{k k^{'}}^{q}( T_{e},T_{l},t)|^{2}S( k,k^{'},T_{e},T_{l},t)\\
 &\times\delta(\varepsilon_{k}-\varepsilon_{k^{'}}+\hbar\omega_{q}( T_{e},T_{l},t)),\\
\end{split}
\end{equation}
where $M_{k k^{'}}^{q}( T_{e},T_{l},t)$  is the electron-phonon  matrix element describing the scattering probability of electrons from initial state at energy $\varepsilon_{k}$ to final state at energy $\varepsilon_{k^{'}}$, and the difference between initial electron energy and final electron energy being equal to the phonon energy $\hbar\omega_{q}$~\cite{allen1987theory}. The indices $k$,\,$k^{'},\,q$ are the initial and final electron wave vector and phonon wave vector, respectively. They are connected via the moment conservation relation ${\bf k}-{\bf k}^{'}={\bf q}$. The thermal factor $S( k,k^{'}, T_{e},T_{l},t)$ has the form
\begin{equation}
 S( k,k^{'}, T_{e},T_{l},t)=[f(\varepsilon_{k},T_{e},t)-f(\varepsilon_{k^{'}},T_{e},t)]n_{B}(\omega_{q},T_{l},t)-f(\varepsilon_{k^{'}},T_{e},t)[1-f(\varepsilon_{k},T_{e},t)],\\
\end{equation}
in which the $f(\varepsilon,T_{e},t)$ are Fermi distribution functions, defined as $f(\varepsilon,T_{e},t)$=$f(\varepsilon,\mu(T_{e}),T_{e},t)$
=${[\exp(\varepsilon-\mu(T_{e}))/K_{B}T_{e}(t)+1]}^{-1}$, where the $\mu(T_{e})$ is the the chemical potential as a function of the electron temperature.
So, the electron-phonon energy exchange rate becomes
\begin{equation}
 \begin{split}
    Z_{ep}(T_{e},T_{l},t)&=\frac{4\pi}{N_{c}\hbar}\sum_{q k}\hbar\omega_{q}(T_{e},T_{l},t)|M_{k k^{'}}^{q}( T_{e},T_{l},t)|^{2}[f(\varepsilon_{k},T_{e},t)-f(\varepsilon_{k^{'}},T_{e},t)]\\
    &\times[n_{B}^{l}(\omega_{q},T_{l},t)-n_{B}^{e}(\omega_{q},T_{e},t)]\delta(\varepsilon_{k}-\varepsilon_{k^{'}}+\hbar\omega_{q}( T_{e},T_{l},t))\,
\end{split}
\label{zep1}
\end{equation}
Here, we introduced the Bose functions for the lattice temperature and the electron temperature $n_B^{l}$ and $n_B^{e}$, respectively. To calculate the formula (\ref{zep1}), the formal procedure is to introduce the Eliashberg function~\cite{allen1987theory}
\begin{equation}
 \begin{split}
    \alpha^{2}F(\varepsilon,\varepsilon^{'},\omega,T_{e},T_{l},t)&=\frac{2}{\hbar N_{c}^{2}g[\mu(T_{e})]}\\
    &\times \sum_{k k^{'}}|M_{k k^{'}}^{q}( T_{e},T_{l},t)|^{2}\delta(\omega-\omega_{q})\delta(\varepsilon-\varepsilon_{k})\delta(\varepsilon^{'}-\varepsilon_{k^{'}})\,.
 \end{split}    
 \label{zep2}
\end{equation}
Here $g[\mu(T_{e})]$ is the $T_{e}$-dependent electron density of states at the chemical potential. Combining the equations (\ref{zep1}) and (\ref{zep2}), we obtain
\begin{equation}
\begin{split}
 Z_{ep}(T_{e},T_{l},t)&=2\pi
 N_{c}g[\mu(T_{e})]\int_{0}^{\infty}\!\int_{-\infty}^{\infty}\!\int_{-\infty}^{\infty}\!
 d\omega d\varepsilon^{'}d\varepsilon
 \;(\hbar\omega)\alpha^{2}F(\varepsilon,\varepsilon^{'},\omega,T_{e},T_{l},t)\\
 &\times[f(\varepsilon,T_{e},t)-f(\varepsilon^{'},T_{e},t)][n_{B}^{l}(\omega,T_{l},t)-n_{B}^{e}(\omega,T_{e},t)]\\
 &\times\delta(\varepsilon-\varepsilon^{'}+\hbar\omega( T_{e},T_{l},t)).
\end{split}
\label{zep3}
\end{equation}
Due to the different energy scales between phonons ($~$meV) and electrons ($~$eV), we adopt some approximations. The first one is for the Eliashberg function, which follows Wang {\em {\em et al.}}~\cite{wang1994time}
\begin{equation}
\begin{split}
   \alpha^{2}F(\varepsilon,\varepsilon^{'},\omega,T_{e},T_{l},t)&=\frac{g(\varepsilon,T_{e},T_{l})g(\varepsilon^{'},T_{e},T_{l})}{g[\mu(T_{e})]}\alpha^{2}F(\mu(T_{e}),\mu(T_{e}),\omega,T_{e},T_{l},t)\\
   &=\frac{g^2(\varepsilon,T_{e},T_{l})}{g[\mu(T_{e})]}\alpha^{2}F(\omega,T_{e},T_{l},t)\,,
\end{split}
\label{zep4}
\end{equation}
in which $g(\varepsilon,T_{e},T_{l}) \approx g(\varepsilon^{'},T_{e},T_{l})$ was used. The second one is for the difference between Fermi distribution function
\begin{equation}
    f(\varepsilon,T_{e})-f(\varepsilon^{'},T_{e})=f(\varepsilon,T_{e})-f(\varepsilon+\hbar\omega,T_{e})=-\hbar\omega\frac{\partial f(\varepsilon,T_{e})}{\partial\varepsilon}.
\label{zep5}
\end{equation}
Inserting Eqs. (\ref{zep4}) \& (\ref{zep5}) into Eq. (\ref{zep3}) yields
\begin{equation}
\begin{split}
     Z_{ep}(T_{e},T_{l},t)&=\frac{2\pi N_{c}}{g[\mu(T_{e})]}\int_{-\infty}^{\infty}g^{2}(\varepsilon,T_{e},T_{l})\frac{\partial f(\varepsilon,T_{e})}{\partial\varepsilon}d\varepsilon\\
     &\times\int_{0}^{\infty}(\hbar\omega)^{2}\alpha^{2}F(\omega,T_{e},T_{l},t)[n_{B}^{e}(\omega,T_{e},t)-n_{B}^{l}(\omega,T_{l},t)]d\omega.\\
\end{split}
\label{zep6}
\end{equation}
From this final expression for electron-phonon energy exchange rate $Z_{ep}(T_{e},T_{l},t)$, the temperature-dependent electron-phonon coupling factor can be obtained by dividing it by the temperature difference between electrons and lattice
\begin{equation}
\begin{split}
        G_{ep}(T_{e},T_{l},t)&=\frac{2\pi N_{c}}{g[\mu(T_{e})](T_{e}-T_{l})}\int_{-\infty}^{\infty}g^{2}(\varepsilon,T_{e},T_{l})\frac{\partial f(\varepsilon,T_{e})}{\partial\varepsilon}d\varepsilon\\
     &\times\int_{0}^{\infty}(\hbar\omega)^{2}\alpha^{2}F(\omega,T_{e},T_{l},t)[n_{B}^{e}(\omega,T_{e},t)-n_{B}^{l}(\omega,T_{l},t)]d\omega.
\end{split}
\label{zep7}
\end{equation}
It should be noted that the final electron-phonon coupling factor is not only a function of electron temperature but also of the lattice temperature. We focus here on a special situation in which the lattice temperature remains at room temperature level. In the high temperature limit($\frac{\hbar\omega}{k_{B}T_{e}}$$\ll$1,$\frac{\hbar\omega}{k_{B}T_{l}}$$\ll$1), the expression for the electron-phonon coupling factor can be reduced to the formula adopted by Lin {\em {\em et al.}}~\cite{lin2008electron}
\begin{equation}
    G_{ep}(T_{e},t)=\frac{N_{c}k_{B}\pi\hbar\lambda\langle w^{2}\rangle}{g(\varepsilon_{F})}\int_{-\infty}^{\infty}g^{2}(\varepsilon)(-\frac{\partial f(\varepsilon,T_{e})}{\partial\varepsilon})d\varepsilon\,,
    \label{zep8}
\end{equation}
where $\lambda\langle w^{2}\rangle=2\int_{0}^{\infty}\omega\alpha^{2}F(\omega)d\omega$ is the second moment of the Eliashberg function and in which the factor $\lambda$ is the electron-phonon coupling constant, $\lambda=2\int_{0}^{\infty}\frac{\alpha^{2}F(\omega)}{\omega}d\omega$. The $T_{e}$-dependent chemical potential $\mu(T_{e})$ can be determined by the equation $N_{e}(T_{e})=\int_{-\infty}^{\infty}g(\varepsilon,T_{e}) f(\varepsilon,\mu(T_{e}),T_{e})d\varepsilon$, in which $N_{e}(T_{e})$ stands for the total number of valence electrons under different excitation~\cite{lin2008electron,bevillon2014free}. In order to calculate this quantity, we use bisection to obtain the root of the corresponding equation. An adaptive gaussian quadrature algorithm is adopted to evaluate all the integrals appearing in the expression (\ref{zep7}). 

\subsection{Calculation  details}
All simulations for obtaining the electron-phonon energy exchange rates were performed using the implementation of density functional theory as given by the open source code ABINIT~\cite{gonze2009abinit,gonze2016recent}. In terms of the electron-phonon coupling, the capabilities in evaluating the electron-phonon matrix element and related properties of phonons are described in Ref.~\cite{gonze1997first,gonze1997dynamical}. The implementation is on the basis of the linear response formalism~\cite{savrasov1996electron}. As it is open source, we were able to modify the code to extract the partial branch-dependent phonon density of state, Eliashberg function and electron-phonon coupling factor. For Aluminium and Copper, we use a norm-conserving electron-ion pseudo-potential under framework of the generalised gradient approximation~\cite{perdew1996generalized}. Three and eleven electrons were treated as valence electrons for aluminium and copper, respectively. The experimental lattice constant for FCC  Aluminium ($4.0496$~\AA) and Copper ($3.61$~\AA) were used. The electron temperature is determined by a Fermi-Dirac distribution with smearing (temperature broadening) ranging from 0.001Ha ($315$ K) to 0.158Ha ($50000$ K). In the calculations of $T_{e}$-dependent electronic density of states, we firstly solve the finite temperature Kohn-Sham equations ~\cite{mermin1965thermal,kohn1965self} to obtain the eigenvalues and then use the tetrahedron method featuring a k-point grid of up to $84\times 84 \times 84$. With increasing electron temperature, we increase the number of bands from 110 to 170 for Aluminium and from 250 to 420 for Copper, respectively. In order to get the $T_{e}$-dependent Eliashberg function, we adopt finite-temperature DFPT method to compute $T_{e}$-dependent electron-phonon matrix elements using the unshifted  k-point grid featuring $32 \times 32 \times 32$ points and a subset thereof for the q-point grid of $8\times 8 \times 8$~\cite{mermin1965thermal,baroni2001phonons}.

\section{Results and discussions}\label{rd}
\subsection{Aluminium}
\begin{figure}[H]
\begin{adjustwidth}{-\extralength}{0cm}
\centering
\includegraphics[width=15cm]{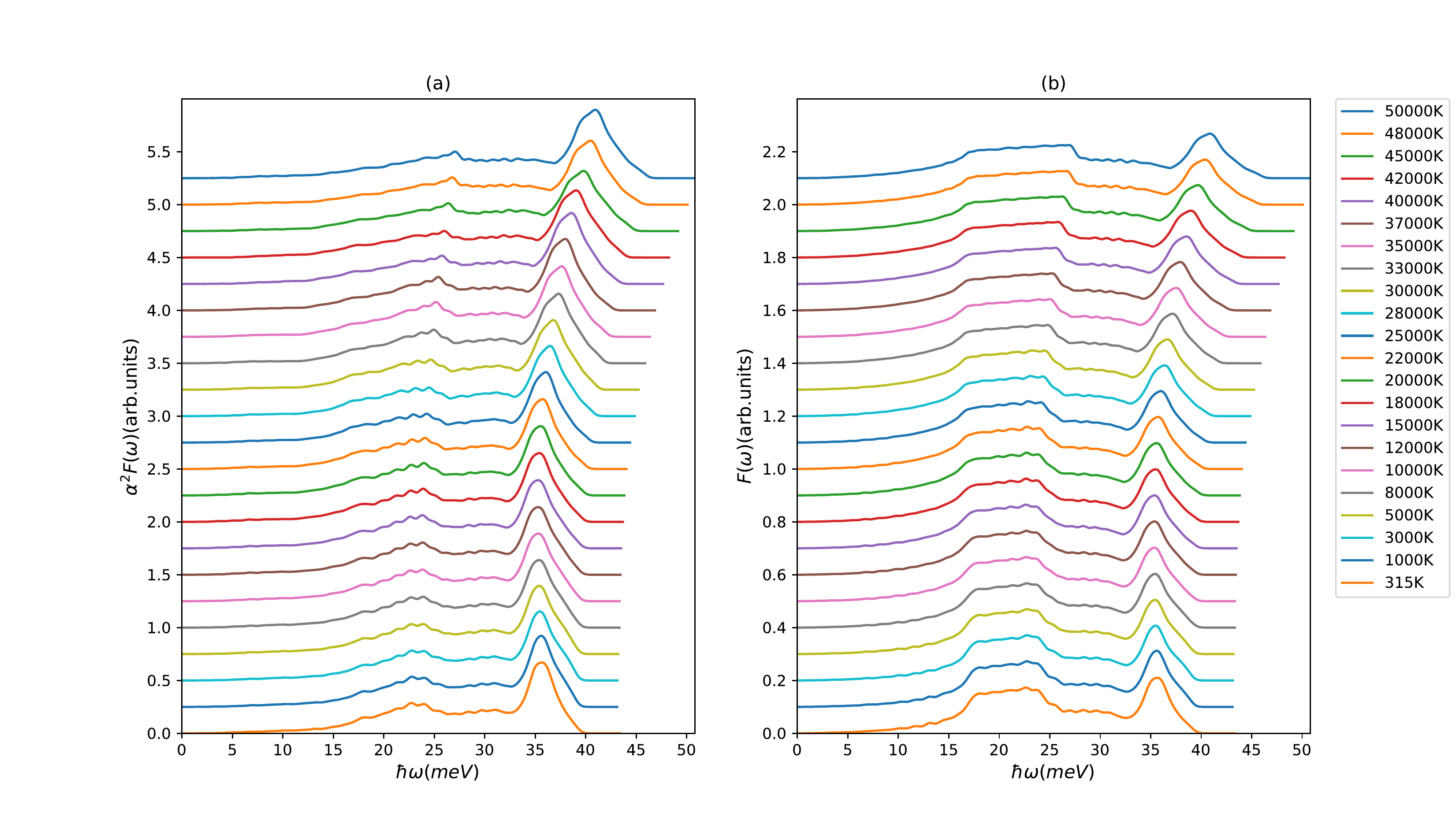}
\end{adjustwidth}
\caption{(a) Eliashberg function $\alpha^{2}F(\omega)$ and (b) phonon density of states $F(\omega)$ of aluminium at different electron temperatures (shifted along the y-axis for better visibility).}
\label{fig:Figure_1_new_al}
\end{figure}
As can be seen from expression (\ref{zep7}), the evaluation of electron-phonon coupling factor is related to the specific phonon states that receive the energy and are determined by the phonon density of states $F(\omega)$ and the electron-phonon coupling as incorporated in the Eliashberg function $\alpha^{2}F(\omega)$. The results for the $T_{e}$-dependent Eliashberg function and phonon DOS of Aluminium are presented in Fig.~\ref{fig:Figure_1_new_al}. We note that the Eliashberg function and the phonon density of states both show continuous and smooth changes with increasing electron temperature. The deviations between $T_e=315$ K and $T_e\sim 20000$ K remain however small. This justifies in this range the often applied approximation of using the ground state Eliashberg function for the energy transfer rate at all electron temperatures. When increasing the electron temperature above $20000$ K, we find that the broadening in the longitudinal peak and the transversal plateau of the phonon density of states as well as the shift to higher frequencies cannot be ignored anymore and needs to be taken into account. As for the corresponding Eliashberg function, similar broadening and shifting of spectral weights can be observed. There is a small redistribution of weight from transversal into the longitudinal channel, we will discuss this in detail below.

\begin{figure}[H]
\begin{adjustwidth}{-\extralength}{0cm}
\centering
\includegraphics[width=15cm]{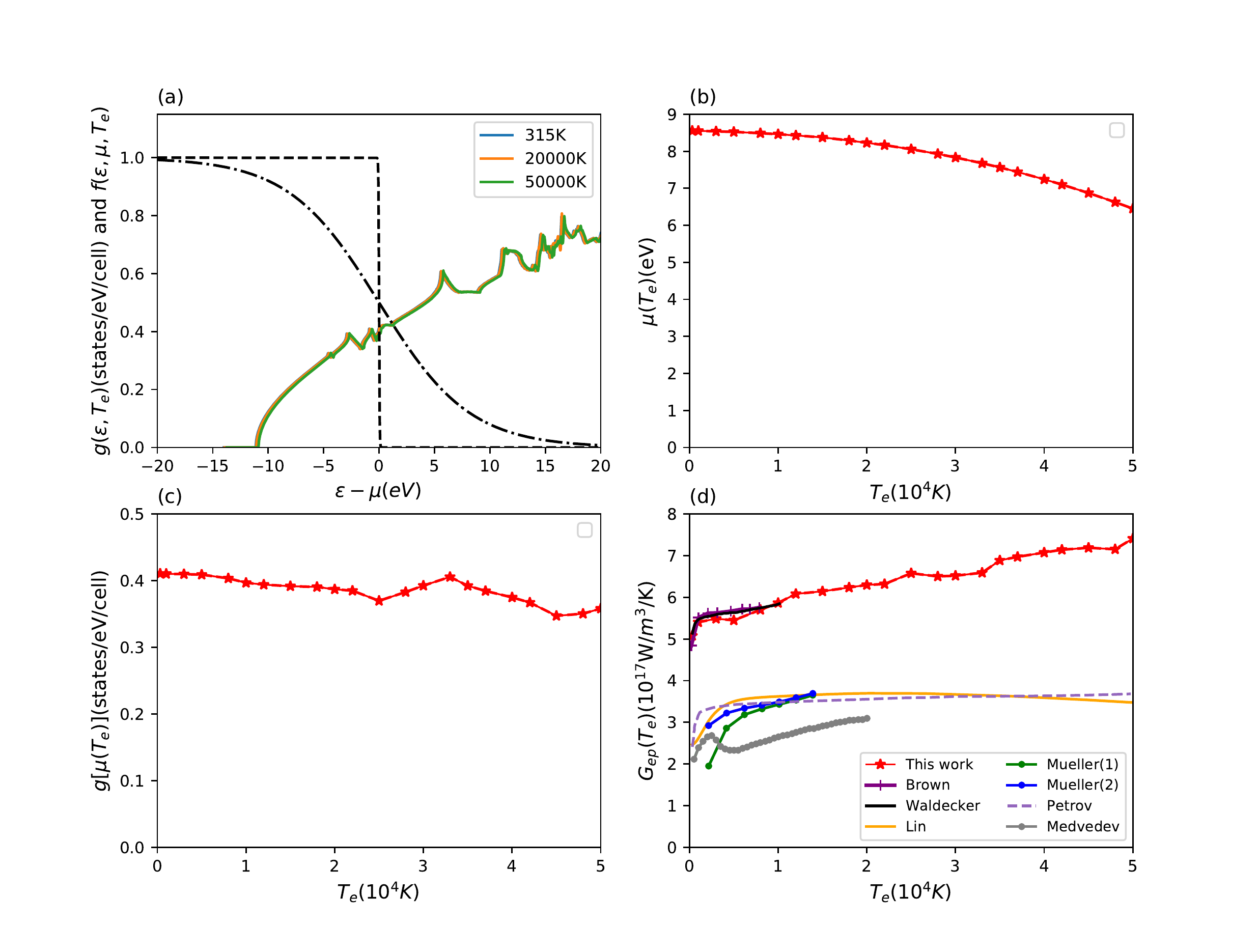}
\end{adjustwidth}
\caption{Results for aluminium.
(a) electron density of states $g(\varepsilon,T_{e})$ with increasing electron temperature and  Fermi distribution function $f(\varepsilon,\mu(T_{e}),T_{e})$ (dashed lines) for two electron temperatures at $315$ K and $50000$ K; (b) chemical potential $\mu(T_{e})$ and (c) electron density of states at chemical potential $g[\mu(T_{e})]$ as a function of electron temperature; (d) electron temperature-dependent electron-phonon coupling factor $ G_{ep}(T_{e})$, compared with various theoretical calculations. The available theoretical data are predicted by Lin {\em et al.}~\cite{lin2008electron}, M\"uller {\em et al.}~\cite{mueller2013relaxation}(their estimations contain two different physical conditions), Petrov {\em et al.}~\cite{petrov2013thermal}, Waldecker {\em et al.}~\cite{waldecker2016electron}, Brown {\em et al.}~\cite{brown2016ab} and Medvedev et
al.~\cite{medvedev2020electron}.}
\label{fig:Figure_2_new_al}
\end{figure}

In addition to the input quantity Eliashberg function, the $T_{e}$-dependent electronic density of states $g(\varepsilon,T_{e})$ and the Fermi distribution function $f(\varepsilon,\mu(T_{e}),T_{e})$ for the occupation numbers of electronic states are important, which determine the contribution of the electronic states around the chemical potential $g[\mu(T_{e})]$ to the energy exchange. The results for the $T_{e}$-dependent electron density of states, chemical potential, electron density of states at the chemical potential and the final electron-phonon coupling factor of Aluminium are shown in Fig. \ref{fig:Figure_2_new_al}. 

From Fig. \ref{fig:Figure_2_new_al}(a), we find that the electron density of states shows little change with increasing electron temperature. The chemical potential decreases for high electron temperatures due to the added contributions from higher energy bands, see Fig. \ref{fig:Figure_2_new_al}(b). Looking at the electron density of states at the chemical potential and the electron-phonon coupling factor, see Fig. \ref{fig:Figure_2_new_al}(c) \& (d), an opposite non-monotonous trend towards high electron temperature can be observed. Especially for electron temperatures between 25000K and 33000K, the chemical potential samples the small features of the DOS around the Fermi edge leading to the structural features of changing slopes in the electron-phonon coupling visible at those temperatures. The overall trend is nevertheless an increase in electron-phonon coupling with temperature.

We compare our results for the electron-phonon coupling to various predictions from different theoretical methods in Fig.~\ref{fig:Figure_2_new_al}(d). Our improved $T_{e}$-dependent calculations are based on our recent work used in Waldecker {\em et al.}~\cite{waldecker2016electron}. They are in good agreement with each other below $20000$K with small deviations due to technical differences in the calculations of Eliashberg function and electron DOS. Brown {\em et al.} adopted a similar DFT based scheme to calculate this quantity and thus their results match well with our estimations in their considered electron temperature range~\cite{brown2016ab}. For the ultrafast interaction of lasers with matter, Lin {\em et al.} used the expression (\ref{zep8}) to investigate the electron-phonon coupling factor under nonequilibrium conditions for a series of metals with different electronic complexity~\cite{lin2008electron}. Contrary to the method used here and by Brown {\em et al.}, their DFT calculations provided only the electron density of states, but the second moment of the Eliashberg function was taken from experiment. Figure~\ref{fig:Figure_2_new_al}(d) shows that the $T_{e}$-dependent electron-phonon coupling factor given by Lin {\em et al.}~\cite{lin2008electron} is smaller by about a factor of two. The reason for this discrepancy was explained by Waldecker {\em et al.} and found to be the inconsistent too early adoption of the two-temperature model in the analysis of experimental data~\cite{waldecker2016electron}.  

Our results show an increasing trend of the electron-phonon coupling factor for high electron temperature. On the contrary, Lin {\em et al.} provide a prediction of a flat or decreasing electron-phonon coupling. Petrov {\em et al.}~\cite{petrov2013thermal} adopted an electron-phonon collision integral method with the effective electron mass for $sp$ electron and the Lindhard approximation, but they only focus on the interaction between electrons and the longitudinal acoustic phonon mode. Their results match Lin {\em et al.} which implies that the adjusting parameter for the second moment of the Eliashberg function used may stand for the partial branch phonon mode instead of the total~\cite{lin2008electron}. M\"uller {\em et al.}~\cite{mueller2013relaxation} adopted a similar method, but in their calculations, they took the free electron approximation and used the jellium model to simplify the formula for the transition matrix element. Their estimations for two different conditions both show a similar trend as Lin {\em et al.} and Petrov {\em et al.}~\cite{lin2008electron,petrov2013thermal}. Medvedev {\em et al.} apply the tight-binding molecular dynamics scheme and their predictions present a similar qualitative trend to our results above $5000$ K electron temperature but at half the magnitude~\cite{medvedev2020electron}. As one can see, empirical calculations seem to underestimate the electron-phonon coupling factor as a function of electron temperature.

\begin{figure}[H]
\begin{adjustwidth}{-\extralength}{0cm}
\centering
\includegraphics[width=15cm]{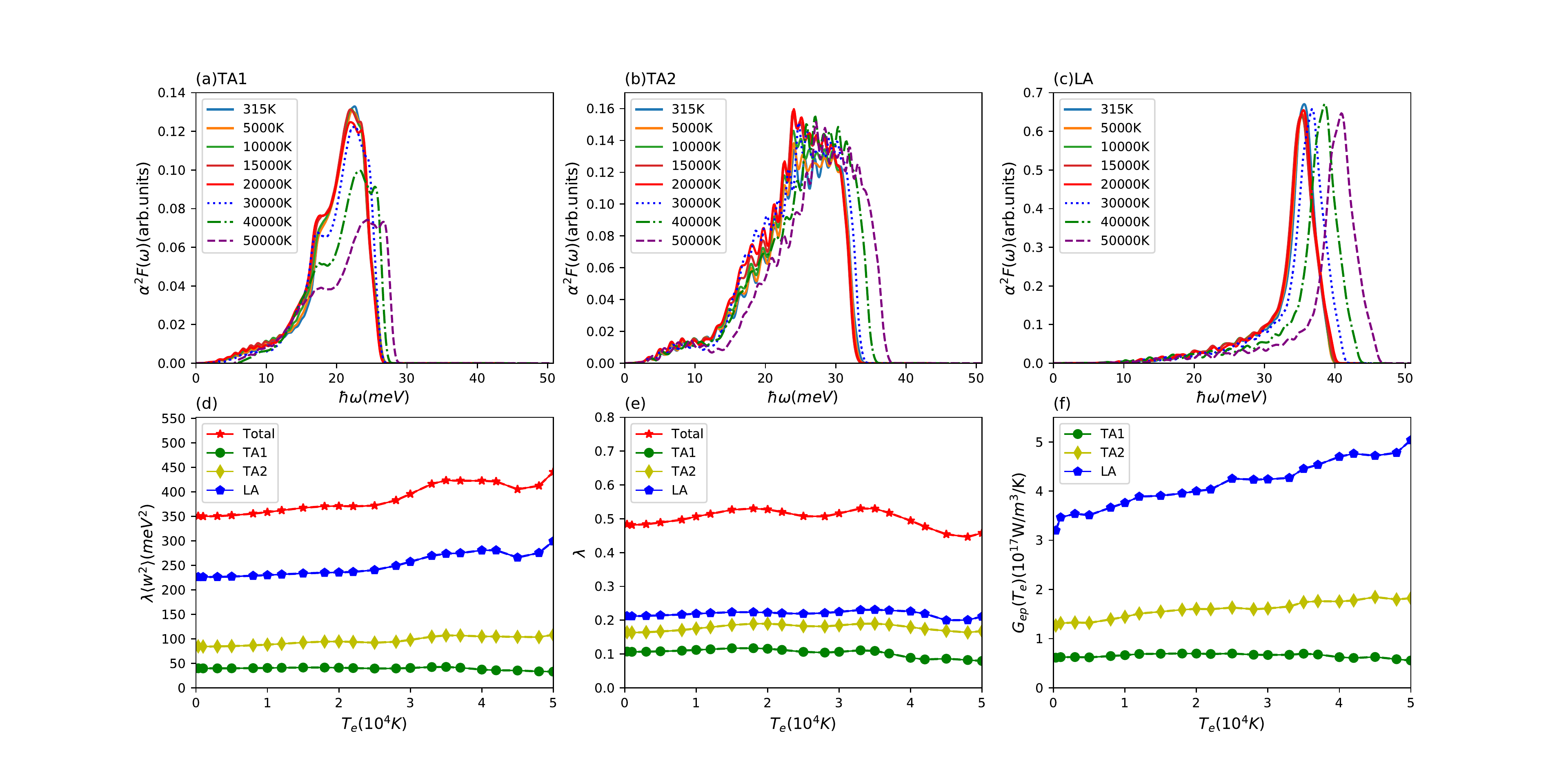}
\end{adjustwidth}
\caption{Results for aluminium. 
Eliashberg function $\alpha^{2}F(\omega)$ of partial branches of (a) TA1, (b) TA2 and (c) LA at different electron temperatures; (d) second moment of Eliashberg function $\lambda\langle w^{2}\rangle$  and (e) electron-phonon coupling constant $\lambda$ for total and three different branches (TA1, TA2, LA) with increasing electron temperature; (f) electron temperature-dependent electron-phonon coupling factor $ G_{ep}(T_{e})$ for three partial branches (TA1, TA2, LA).}
\label{fig:Figure_3_new_al}
\end{figure}

In order to give a deeper insight into the energy transfer channels between electron and phonon subsystems, we divide the phonon subsystem into three individual parts and compute the $T_{e}$-dependent scattering matrix elements corresponding to the three different acoustic modes. Figure~\ref{fig:Figure_3_new_al}(a)-(c) presents the partial Eliashberg functions for the two acoustic (TA1, TA2) and one longitudinal (LA) branch at different electron temperatures. We find that these three partial Eliashberg functions do not change much below $20000$ K as was observed above for the total Eliashberg function. With further increasing electron temperature, the peaks all move to higher frequencies. The amplitude of the TA1 Eliashberg function decreases but the other two partial Eliashberg function's amplitude for TA2 and LA remain basically unchanged. Using these partial functions, we obtained the partial $T_{e}$-dependent electron phonon coupling factors, which are displayed in Fig. \ref{fig:Figure_3_new_al}(f). We can see that these three partial electron-phonon coupling factor have a big difference on the quantitative level and the longitudinal acoustic mode plays a dominant role in the  electron-phonon coupling. It indicates that the phonon subsystem is likely to undergo a nonequilibrium energy relaxation dynamics. Also, it becomes clear that the increase of electron-phonon coupling with electron temperature stems from the increased coupling of the electrons to the longitudinal mode. From the total and partial Eliashberg function, we also computed the second moment of Eliashberg function and electron-phonon coupling constant, which is linked to the conventional superconductivity critical temperature $T_{c}$. The results are presented in Figs. \ref{fig:Figure_3_new_al}(d) \& (e). We can see that the second moment of the total and partial Eliashberg function towards high electron temperature has a similar trend to the $T_{e}$-dependent total and partial electron-phonon coupling factor. In our case,the total electron phonon coupling constant at $T_{e}$=315K is $\lambda=0.48$, which is in good agreement with our recent work~\cite{waldecker2016electron}, other DFT prediction~\cite{savrasov1996electron}, and gives a reasonable critical superconductivity temperature within the McMillan model for Al of $T_c=1.48$~K~\cite{Allen1975}. As mentioned above, the longitudinal $\lambda\langle\omega^2\rangle=226meV^2$ is close the value that Lin {\em et al.} used for the total electron-phonon coupling factor.

\subsection{Copper}
\begin{figure}[H]
\begin{adjustwidth}{-\extralength}{0cm}
\centering
\includegraphics[width=15cm]{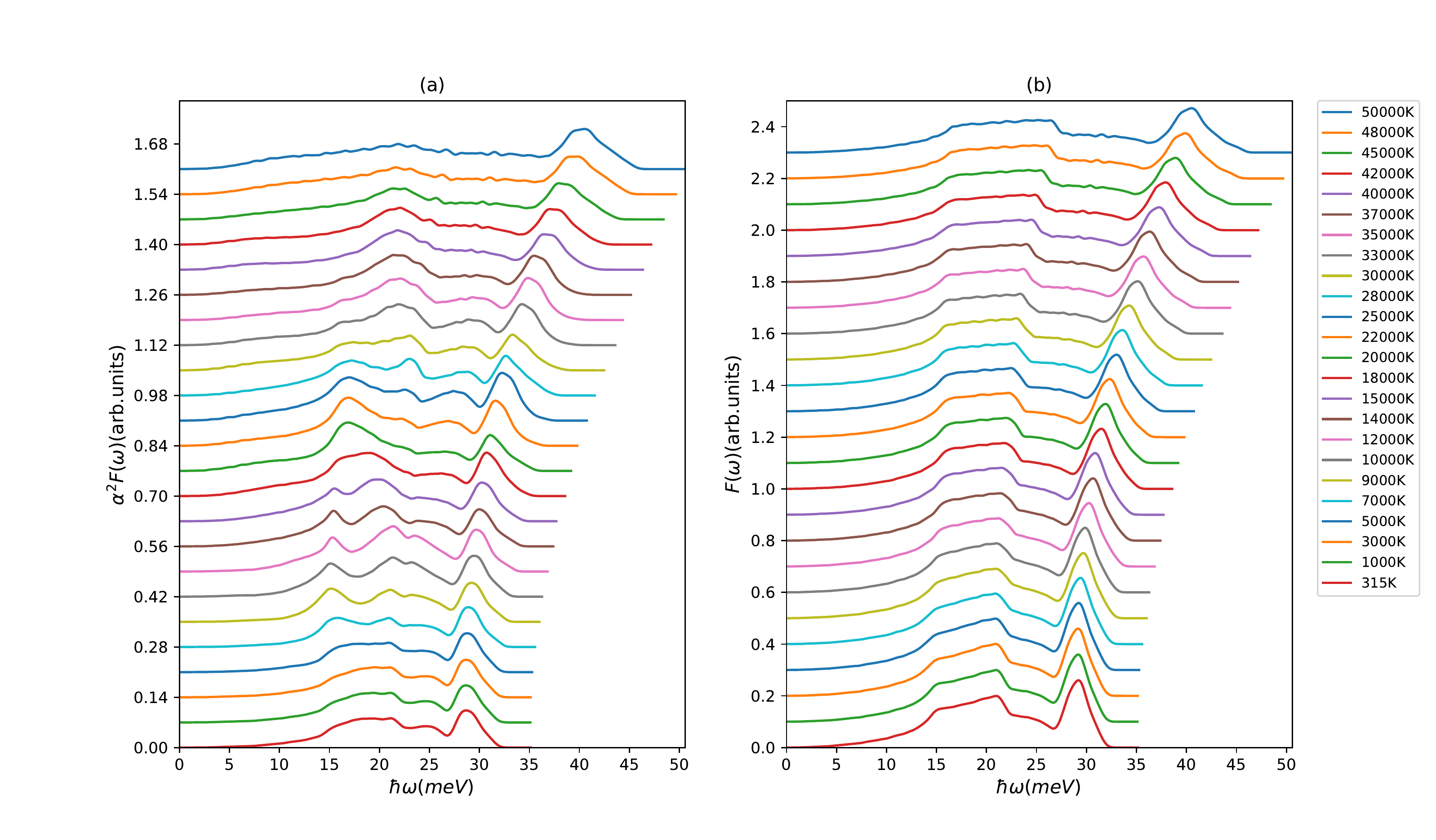}
\end{adjustwidth}
\caption{(a) Eliashberg function $\alpha^{2}F(\omega)$ and (b) phonon density of states $F(\omega)$ of copper at different electron temperatures (shifted along the y-axis for better visibility).}
\label{fig:Figure_1_new_cu}
\end{figure}

The case of the transition metal copper is, compared to the situation in the simple metal Aluminium, slightly more involved. From Fig. \ref{fig:Figure_1_new_cu}(a), it is obvious that the phonon density of states towards high electron temperature shows similar behaviour as presented in the Aluminium case. The longitudinal peak shifts monotonously to higher frequencies and broadens at the same time. The plateau at $20$ meV stays around this energy but broadens considerably. These are consistent with the behaviour of the phonon DOS found in our previous
work~\cite{zhang2018lattice}. However, the changes within the Eliashberg function at different electron temperature is more complicated especially between 10000K and 30000K. Whereas the longitudinal mode exhibits the now well-known shift and broadening, the maxima stemming from the transversal modes shift in magnitude several times until at the highest temperatures almost no feature of them is left.

\begin{figure}[H]
\begin{adjustwidth}{-\extralength}{0cm}
\centering
\includegraphics[width=15cm]{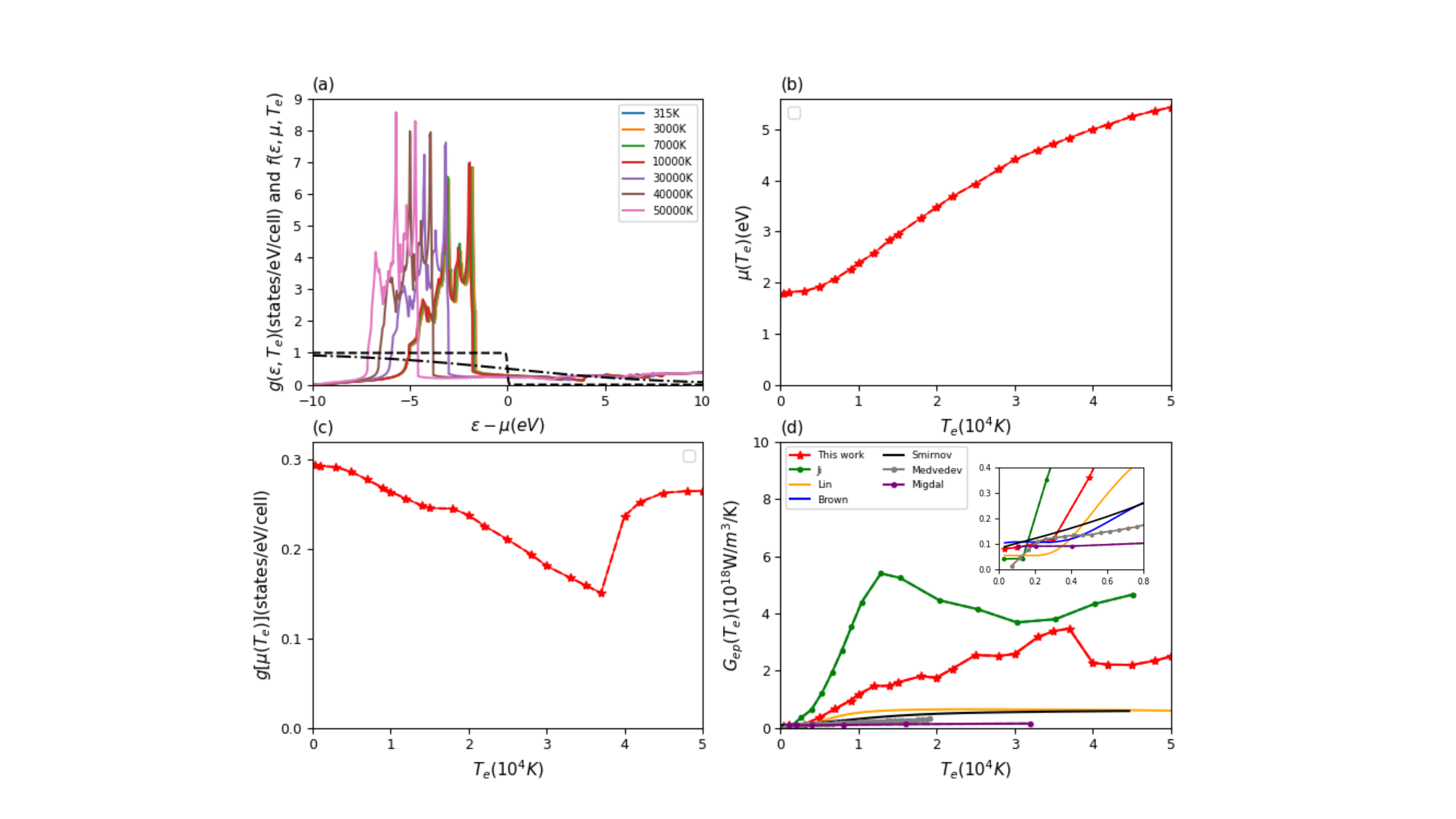}
\end{adjustwidth}
\caption{Results for copper.
(a) electron density of states $g(\varepsilon,T_{e})$ with increasing electron temperature and  Fermi distribution function $f(\varepsilon,\mu(T_{e}),T_{e})$ (dashed lines) for two electron temperatures at $315$ K and $50000$ K; (b) chemical potential $\mu(T_{e})$ and (c) electron density of states at chemical potential $g[\mu(T_{e})]$ as a function of electron temperature; (d) electron temperature-dependent electron-phonon coupling factor $ G_{ep}(T_{e})$, compared with various theoretical calculations.The available theoretical data are estimated by Lin {\em et al.}~\cite{lin2008electron}, Migdal {\em et al.}~\cite{migdal2016heat}, Ji {\em et al.}~\cite{ji2016ab}, Brown {\em et al.}~\cite{brown2016ab}, Smirmov {\em et al.}~\cite{smirnov2020copper} and Medvedev {\em et al.}~\cite{medvedev2020electron}.}
\label{fig:Figure_2_new_cu}
\end{figure}
Fig. \ref{fig:Figure_2_new_cu}(a) shows the electron DOS of copper with typical features for a d-row metal. We can see that the electron DOS shows little changes under low electronic excitations below 10000K. For higher electronic excitations, the electronic structure of copper undergoes dramatic changes from 10000K to 50000K. For the highest electron temperature considered here, these changes have brought about that (i) all maxima are higher, (ii) the left maximum is now the highest, (iii) the DOS is compressed in energy range. With the increasing electron temperature, the Fermi energy shifts to the right. It means that the chemical potential will increase up to high electron temperatures, as can be seen in the Fig.~\ref{fig:Figure_2_new_cu}(b). Fig.~\ref{fig:Figure_2_new_cu}(c) shows the electron DOS at the chemical potential. There is a decrease below $37000$~K but from $37000$~K onward the trend reverses and the value increases. Thus, our electron-phonon coupling factor presented in Fig. \ref{fig:Figure_2_new_cu}(d) shows a characteristic dip at $37000$~K due to the jump in the electron DOS at the chemical potential.

The room temperature ($T_{e}=315$ K) value is $8.31\times10^{16}\,W/m^{3}/K$ and agrees well with other DFT calculations implemented by Brown {\em et al.}~\cite{brown2016ab} and Smirnov {\em et al.}~\cite{smirnov2020copper}. As for the variation of the electron-phonon coupling factor with electron temperature, Ji {\em et al.} shows a similar qualitative trend towards high electron temperature but the actual values are twice as high~\cite{ji2016ab}. Although they too adopted a finite temperature DFT based scheme, they used expression (\ref{zep8}) instead of expression (\ref{zep7}). In our expression (\ref{zep7}), we used the $T_{e}$-dependent electron DOS at chemical potential instead of $T_{e}$-dependent electron DOS at Fermi energy and we don't take the high temperature approximation. Further, our Eliashberg functions at elevated temperatures do not agree with Ji {\em et al.}. 

The other results for the electron-phonon coupling shown in Fig.~\ref{fig:Figure_2_new_cu} group together at half the magnitude. 
Lin {\em et al.} and Smirnov both give very similar predictions with little increase for higher temperatures, even though some improvements have been made on the Eliashberg function and the used formula in the case of Smirnov~\cite{lin2008electron,smirnov2020copper}. Their underestimation may be attributed to the fact that they don't take the electronic excitation effect on electron DOS and Eliashberg function into account. The trend predicted by Brown {\em et al.} shows good agreement with Smirnov {\em et al.}~\cite{smirnov2020copper,brown2016ab}. Medvedev {\em et al.} and Migdal {\em et al.} both present an increasing trend but provide an underestimation possibly due to the empirical calculation scheme~\cite{medvedev2020electron,migdal2016heat}. In general, the case of d-row elements is a non-trivial one as tiny changes in the very spiky DOS have a large influence on the electron-phonon coupling.

\begin{figure}[H]
\begin{adjustwidth}{-\extralength}{0cm}
\centering
\includegraphics[width=15cm]{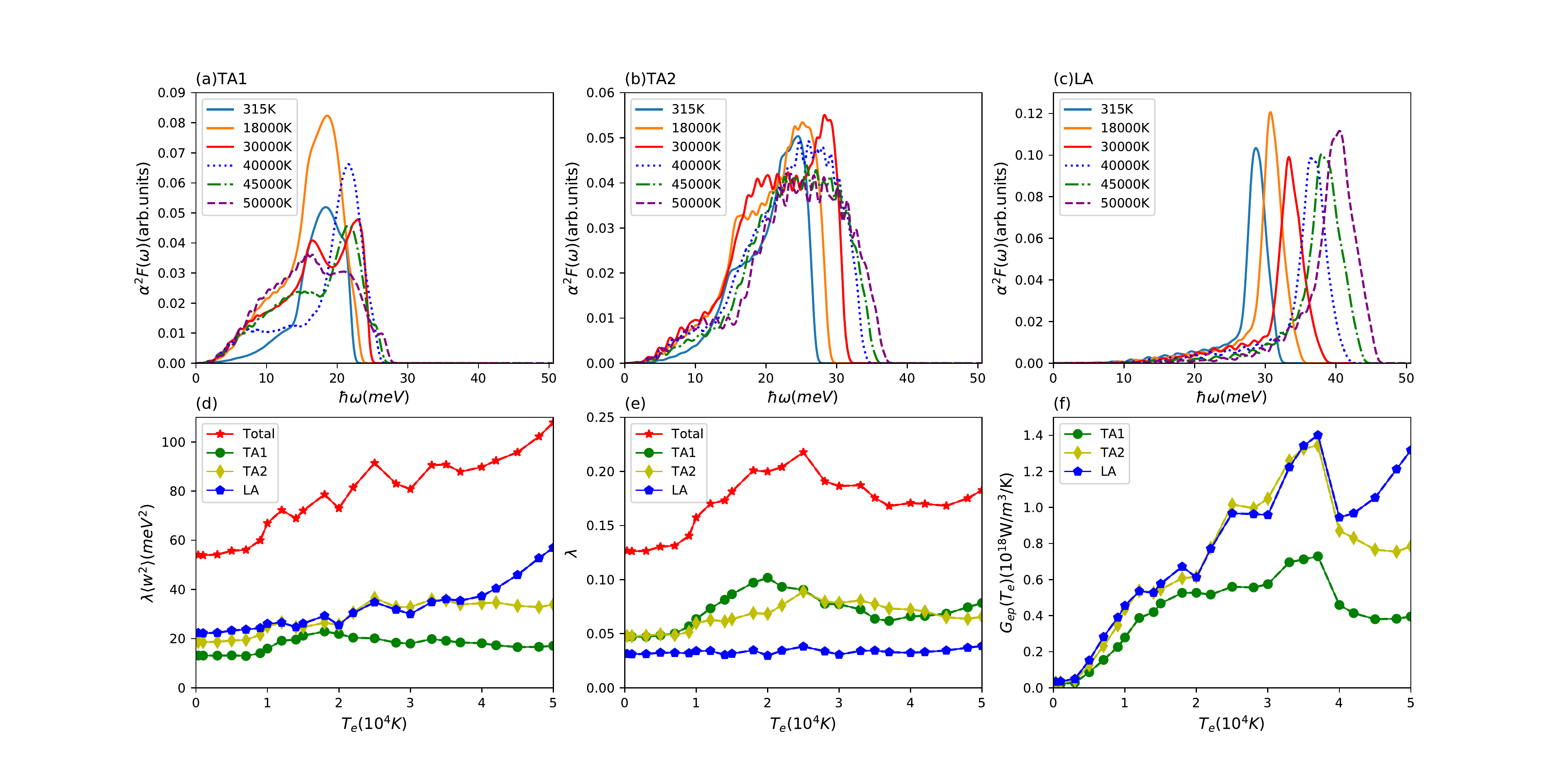}
\end{adjustwidth}
\caption{Results for copper. 
Eliashberg function $\alpha^{2}F(\omega)$ of partial branches of (a) TA1, (b) TA2 and (c) LA at different electron temperatures; (d) second moment of Eliashberg function $\lambda\langle w^{2}\rangle$  and (e) electron-phonon coupling constant $\lambda$ for total and three different branches (TA1, TA2, LA) with increasing electron temperature; (f) electron temperature-dependent electron-phonon coupling factor $ G_{ep}(T_{e})$ for three partial branches (TA1, TA2, LA).}
\label{fig:Figure_3_new_cu}
\end{figure}
The complex behaviour of the total Eliashberg function in the electron temperature range from $10000$ K to $30000$ K, is reflected by the trends of the partial Eliashberg functions presented in Fig.~\ref{fig:Figure_3_new_cu}(a), (b) \& (c). Whereas the longitudinal part shows a smooth shift to higher energies with temperature and almost no change in magnitude and the TA2 mode only shows a broadening, the TA1 mode displays several non-trivial changes. Interestingly, this most complex variation of the TA1 Eliashberg function is not reflected in the oscillations of the second moment of the Eliashberg function. Here, in particular the TA2 and LA modes show structure in the corresponding electron temperature range, as seen from Fig. \ref{fig:Figure_3_new_cu}(d). 

In our present work, the total second moment of the Eliashberg function $\lambda\langle w^{2}\rangle$  and electron-phonon coupling constant $\lambda$, see Fig.~\ref{fig:Figure_3_new_cu}(d) \& (e), at $T_{e}=315$ K are $54\,meV^{2}$ and $0.13$ respectively, which are in good agreement with Ji {\em et al.}~\cite{ji2016ab} and other first principles calculations~\cite{savrasov1996electron} as well as experimentally extracted value by Obergfell {\em et al.}~\cite{obergfell2020tracking}. Above $40000$ K, the partial second moment of the Eliashberg function for the longitudinal and transversal modes show diverging tendency with the longitudinal mode starting to dominate. The same trend is seen in the partial electron-phonon coupling factors in Fig.~\ref{fig:Figure_3_new_cu}(f). Thus, it can be expected that for electron temperatures below $\sim 30000$~K, where all partial electron-phonon couplings are of very similar size, a two-temperature model is sufficient, but for higher electron temperatures an improved model similar to the aluminium case might be needed~\cite{waldecker2016electron}. In this case, of course also phonon-phonon coupling needs to be investigated~\cite{sadasivam2017theory,maldonado2017theory,ono2018thermalization,ritzmann2020theory}.


\section{Conclusions}
We have studied the energy transfer rate between electrons and phonons under non-equilibrium conditions that occur in ultrafast-laser irradiated aluminium and copper. We used first-principles calculations based on finite temperature DFT and DFPT to accurately determine the $T_{e}$-dependent total and corresponding partial electron-phonon coupling factors $G_{ep}(T_{e})$. In particular we took care to calculate the electron DOS and the Eliashberg function for all considered electron temperatures. Thus, we were able to show for which electron temperatures the often used $T=0$-approximation for the Eliashberg function works.

In the case of aluminium, consistent calculations which obtain all input quantities from first principles seem to agree. More approximate theories or theories that take some input from experiment give different (lower) results for the electron-phonon coupling. This discrepancy stems from the peculiarly dominant role of the longitudinal phonon mode in aluminium which breaks the approximations inherent in the two-temperature model and facilitates the need for a better model featuring several different phonon temperatures. The lower electron-phonon energy transfer rates can be matched well when only considering the energy transfer via the longitudinal mode.

The case of copper is more complicated. Even though the lattice symmetry is the same as for aluminium, the three branches of the phonons contribute equally to the electron-phonon energy transfer. Thus, it can be expected that a two-temperature model is a better approximation for copper than it is for aluminium. However, due to copper being a d-band metal, the changes in the electron DOS, the phonon DOS, and the Eliashberg function with electron temperature are less trivial than for aluminium and need to be fully taken into account. In particular the spiky structure of the electron DOS that gets sampled to varying degrees for increasing electron temperatures is a cause for small-scale variations in the energy transfer rate. 
The importance of the accuracy of the electron DOS for transition metals cannot be understated as the DOS at the chemical potential can have a huge influence on the energy transfer rate. Overcoming this problem will require to calculate the fully electron-energy resolved Eliashberg function.


In this work, we mainly focus on the energy relaxation in the initial excited stage when the lattice temperature remains cold. Should the ion temperature rise to values close to melting and above due to the electron-phonon coupling, the whole excited system will enter a transient and nonequilibrium warm dense matter (WDM) state. In this exotic state, the existing nonlinear and strong coupling effects due to the rising ion temperature will complicate the determination of the electron-ion coupling factor $G_{ei}(T_{e},T_{i})$. Owing to this situation which is highly related to our case, we hope to extend our scheme to cover this interesting and open problem. 

The ultrafast melting upon laser irradiation remains poorly understood even though many advance have been made~\cite{mo2018heterogeneous,smirnov2020copper,molina2021molecular}. In order to model non-equilibrium lattice dynamics on the atomic level, a possible way is to perform two-temperature molecular dynamics (2TMD) simulations~\cite{ivanov2003combined}. From our ab initio results for the partial electron-phonon coupling factors, we conclude that modifications of the Langevin dynamics for the ions are necessary to include differently heated phonon modes~\cite{tamm2018langevin}. We also want to point out that the melting temperature is not constant but a function of the electron temperature due to the electronic excitation effect~\cite{recoules2006effect,yan2016different,zhang2018lattice,smirnov2020copper}. Combining our new values for the electron-phonon coupling factor with the enhanced 2TMD framework, should further increase our understanding of the melting under the non-equilibrium conditions.

\acknowledgments{Jia Zhang thanks China Scholarship Council for financial support. We gratefully acknowledge the CPU time on the hypnos and hemera clusters at the High Performance Computing at HZDR and on a Bull Cluster at the Center for Information Services and High Performance Computing (ZIH) at Technische Universit\"at Dresden.}

\bibliography{reference.bib}

\begin{thebibliography}{999}

\bibitem[Macchi(2013)]{Macchi:2013}
Macchi, A.
\newblock {\em A Superintense Laser-Plasma Interaction Theory Primer};
  Springer: Dordrecht,  2013.

\bibitem[Gamaly and Rode(2013)]{Gamaly:2013}
Gamaly, E.; Rode, A.
\newblock Physics of ultra-short laser interaction with matter: From phonon
  excitation to ultimate transformations.
\newblock {\em Progress in Quantum Electronics} {\bf 2013}, {\em 37},~215--323.
\newblock
  doi:{\changeurlcolor{black}\href{https://doi.org/https://doi.org/10.1016/j.pquantelec.2013.05.001}{\detokenize{https://doi.org/10.1016/j.pquantelec.2013.05.001}}}.

\bibitem[Waldecker \em{et~al.}(2015)Waldecker, Bertoni, and
  Ernstorfer]{waldecker2015compact}
Waldecker, L.; Bertoni, R.; Ernstorfer, R.
\newblock Compact femtosecond electron diffractometer with 100 keV electron
  bunches approaching the single-electron pulse duration limit.
\newblock {\em Journal of Applied Physics} {\bf 2015}, {\em 117},~044903.

\bibitem[Bostedt \em{et~al.}(2016)Bostedt, Boutet, Fritz, Huang, Lee, Lemke,
  Robert, Schlotter, Turner, and Williams]{bostedt2016linac}
Bostedt, C.; Boutet, S.; Fritz, D.M.; Huang, Z.; Lee, H.J.; Lemke, H.T.;
  Robert, A.; Schlotter, W.F.; Turner, J.J.; Williams, G.J.
\newblock Linac coherent light source: The first five years.
\newblock {\em Reviews of Modern Physics} {\bf 2016}, {\em 88},~015007.

\bibitem[Seddon \em{et~al.}(2017)Seddon, Clarke, Dunning, Masciovecchio, Milne,
  Parmigiani, Rugg, Spence, Thompson, Ueda, et~al.]{seddon2017short}
Seddon, E.; Clarke, J.; Dunning, D.; Masciovecchio, C.; Milne, C.; Parmigiani,
  F.; Rugg, D.; Spence, J.; Thompson, N.; Ueda, K.;  et~al.
\newblock Short-wavelength free-electron laser sources and science: a review.
\newblock {\em Reports on Progress in Physics} {\bf 2017}, {\em 80},~115901.

\bibitem[Hofherr \em{et~al.}(2020)Hofherr, Häuser, Dewhurst, Tengdin,
  Sakshath, Nembach, Weber, Shaw, Silva, Kapteyn, Cinchetti, Rethfeld, Murnane,
  Steil, Stadtmüller, Sharma, Aeschlimann, and Mathias]{Hofherr:2020}
Hofherr, M.; Häuser, S.; Dewhurst, J.K.; Tengdin, P.; Sakshath, S.; Nembach,
  H.T.; Weber, S.T.; Shaw, J.M.; Silva, T.J.; Kapteyn, H.C.;  et~al.
\newblock Ultrafast optically induced spin transfer in ferromagnetic alloys.
\newblock {\em Science Advances} {\bf 2020}, {\em 6},~eaay8717,
  \href{http://xxx.lanl.gov/abs/https://www.science.org/doi/pdf/10.1126/sciadv.aay8717}{{\normalfont
  [https://www.science.org/doi/pdf/10.1126/sciadv.aay8717]}}.
\newblock
  doi:{\changeurlcolor{black}\href{https://doi.org/10.1126/sciadv.aay8717}{\detokenize{10.1126/sciadv.aay8717}}}.

\bibitem[Ng(2012)]{ng2012outstanding}
Ng, A.
\newblock Outstanding questions in electron--ion energy relaxation, lattice
  stability, and dielectric function of warm dense matter.
\newblock {\em International Journal of Quantum Chemistry} {\bf 2012}, {\em
  112},~150--160.

\bibitem[Vorberger \em{et~al.}(2010)Vorberger, Gericke, Bornath, and
  Schlanges]{Vorberger:2010}
Vorberger, J.; Gericke, D.O.; Bornath, T.; Schlanges, M.
\newblock Energy relaxation in dense, strongly coupled two-temperature plasmas.
\newblock {\em Phys. Rev. E} {\bf 2010}, {\em 81},~046404.
\newblock
  doi:{\changeurlcolor{black}\href{https://doi.org/10.1103/PhysRevE.81.046404}{\detokenize{10.1103/PhysRevE.81.046404}}}.

\bibitem[Dorchies and Recoules(2016)]{dorchies2016non}
Dorchies, F.; Recoules, V.
\newblock Non-equilibrium solid-to-plasma transition dynamics using XANES
  diagnostic.
\newblock {\em Physics Reports} {\bf 2016}, {\em 657},~1--26.

\bibitem[Rethfeld \em{et~al.}(2017)Rethfeld, Ivanov, Garcia, and
  Anisimov]{rethfeld2017modelling}
Rethfeld, B.; Ivanov, D.S.; Garcia, M.E.; Anisimov, S.I.
\newblock Modelling ultrafast laser ablation.
\newblock {\em Journal of Physics D: Applied Physics} {\bf 2017}, {\em
  50},~193001.

\bibitem[Rethfeld \em{et~al.}(2002)Rethfeld, Kaiser, Vicanek, and
  Simon]{rethfeld2002ultrafast}
Rethfeld, B.; Kaiser, A.; Vicanek, M.; Simon, G.
\newblock Ultrafast dynamics of nonequilibrium electrons in metals under
  femtosecond laser irradiation.
\newblock {\em Physical Review B} {\bf 2002}, {\em 65},~214303.

\bibitem[Silaeva \em{et~al.}(2018)Silaeva, B{\'e}villon, Stoian, and
  Colombier]{silaeva2018ultrafast}
Silaeva, E.; B{\'e}villon, E.; Stoian, R.; Colombier, J.P.
\newblock Ultrafast electron dynamics and orbital-dependent thermalization in
  photoexcited metals.
\newblock {\em Physical Review B} {\bf 2018}, {\em 98},~094306.

\bibitem[Weber and Rethfeld(2019)]{weber2019phonon}
Weber, S.T.; Rethfeld, B.
\newblock Phonon-induced long-lasting nonequilibrium in the electron system of
  a laser-excited solid.
\newblock {\em Physical Review B} {\bf 2019}, {\em 99},~174314.

\bibitem[Obergfell and Demsar(2020)]{obergfell2020tracking}
Obergfell, M.; Demsar, J.
\newblock Tracking the time evolution of the electron distribution function in
  copper by femtosecond broadband optical spectroscopy.
\newblock {\em Physical review letters} {\bf 2020}, {\em 124},~037401.

\bibitem[Lee \em{et~al.}(2021)Lee, Kim, Kang, Vinko, Bae, Cho, Chung, Kim,
  Kwon, Lee, et~al.]{lee2021investigation}
Lee, J.W.; Kim, M.; Kang, G.; Vinko, S.M.; Bae, L.; Cho, M.S.; Chung, H.K.;
  Kim, M.; Kwon, S.; Lee, G.;  et~al.
\newblock Investigation of Nonequilibrium Electronic Dynamics of Warm Dense
  Copper with Femtosecond X-Ray Absorption Spectroscopy.
\newblock {\em Physical Review Letters} {\bf 2021}, {\em 127},~175003.

\bibitem[Chen \em{et~al.}(2021)Chen, Tsui, Mo, Fedosejevs, Ozaki, Recoules,
  Sterne, and Ng]{chen2021electron}
Chen, Z.; Tsui, Y.; Mo, M.; Fedosejevs, R.; Ozaki, T.; Recoules, V.; Sterne,
  P.; Ng, A.
\newblock Electron Kinetics Induced by Ultrafast Photoexcitation of Warm Dense
  Matter in a 30-nm-Thick Foil.
\newblock {\em Physical Review Letters} {\bf 2021}, {\em 127},~097403.

\bibitem[Grolleau \em{et~al.}(2021)Grolleau, Dorchies, Jourdain, Phuoc,
  Gautier, Mahieu, Renaudin, Recoules, Martinez, and
  Lecherbourg]{grolleau2021femtosecond}
Grolleau, A.; Dorchies, F.; Jourdain, N.; Phuoc, K.T.; Gautier, J.; Mahieu, B.;
  Renaudin, P.; Recoules, V.; Martinez, P.; Lecherbourg, L.
\newblock Femtosecond Resolution of the Nonballistic Electron Energy Transport
  in Warm Dense Copper.
\newblock {\em Physical Review Letters} {\bf 2021}, {\em 127},~275901.

\bibitem[Ono(2017)]{ono2017nonequilibrium}
Ono, S.
\newblock Nonequilibrium phonon dynamics beyond the quasiequilibrium approach.
\newblock {\em Physical Review B} {\bf 2017}, {\em 96},~024301.

\bibitem[Klett and Rethfeld(2018)]{klett2018relaxation}
Klett, I.; Rethfeld, B.
\newblock Relaxation of a nonequilibrium phonon distribution induced by
  femtosecond laser irradiation.
\newblock {\em Physical Review B} {\bf 2018}, {\em 98},~144306.

\bibitem[Recoules \em{et~al.}(2006)Recoules, Cl{\'e}rouin, Z{\'e}rah, Anglade,
  and Mazevet]{recoules2006effect}
Recoules, V.; Cl{\'e}rouin, J.; Z{\'e}rah, G.; Anglade, P.; Mazevet, S.
\newblock Effect of intense laser irradiation on the lattice stability of
  semiconductors and metals.
\newblock {\em Physical review letters} {\bf 2006}, {\em 96},~055503.

\bibitem[Yan \em{et~al.}(2016)Yan, Cheng, Zhang, Zhu, and
  Ren]{yan2016different}
Yan, G.Q.; Cheng, X.L.; Zhang, H.; Zhu, Z.Y.; Ren, D.H.
\newblock Different effects of electronic excitation on metals and
  semiconductors.
\newblock {\em Physical Review B} {\bf 2016}, {\em 93},~214302.

\bibitem[Zhang \em{et~al.}(2018)Zhang, Cheng, He, and Yan]{zhang2018lattice}
Zhang, J.; Cheng, X.; He, N.; Yan, G.
\newblock Lattice response to the relaxation of electronic pressure of
  ultrafast laser-irradiated copper and nickel nanofilms.
\newblock {\em Journal of Physics: Condensed Matter} {\bf 2018}, {\em
  30},~085401.

\bibitem[Ben-Mahfoud \em{et~al.}(2021)Ben-Mahfoud, Silaeva, Stoian, and
  Colombier]{ben2021structural}
Ben-Mahfoud, L.; Silaeva, E.; Stoian, R.; Colombier, J.P.
\newblock Structural instability of transition metals upon ultrafast laser
  irradiation.
\newblock {\em Physical Review B} {\bf 2021}, {\em 104},~104104.

\bibitem[Ono and Kobayashi(2021)]{ono2021lattice}
Ono, S.; Kobayashi, D.
\newblock Lattice stability of ordered Au-Cu alloys in the warm dense matter
  regime.
\newblock {\em Physical Review B} {\bf 2021}, {\em 103},~094114.

\bibitem[Harb \em{et~al.}(2008)Harb, Ernstorfer, Hebeisen, Sciaini, Peng,
  Dartigalongue, Eriksson, Lagally, Kruglik, and
  Miller]{harb2008electronically}
Harb, M.; Ernstorfer, R.; Hebeisen, C.T.; Sciaini, G.; Peng, W.; Dartigalongue,
  T.; Eriksson, M.A.; Lagally, M.G.; Kruglik, S.G.; Miller, R.D.
\newblock Electronically driven structure changes of Si captured by femtosecond
  electron diffraction.
\newblock {\em Physical review letters} {\bf 2008}, {\em 100},~155504.

\bibitem[Giret \em{et~al.}(2014)Giret, Daraszewicz, Duffy, Shluger, and
  Tanimura]{giret2014nonthermal}
Giret, Y.; Daraszewicz, S.L.; Duffy, D.M.; Shluger, A.L.; Tanimura, K.
\newblock Nonthermal solid-to-solid phase transitions in tungsten.
\newblock {\em Physical Review B} {\bf 2014}, {\em 90},~094103.

\bibitem[Lian \em{et~al.}(2016)Lian, Zhang, and Meng]{lian2016ab}
Lian, C.; Zhang, S.; Meng, S.
\newblock Ab initio evidence for nonthermal characteristics in ultrafast laser
  melting.
\newblock {\em Physical Review B} {\bf 2016}, {\em 94},~184310.

\bibitem[Mo \em{et~al.}(2018)Mo, Chen, Li, Dunning, Witte, Baldwin, Fletcher,
  Kim, Ng, Redmer, et~al.]{mo2018heterogeneous}
Mo, M.; Chen, Z.; Li, R.; Dunning, M.; Witte, B.; Baldwin, J.; Fletcher, L.;
  Kim, J.; Ng, A.; Redmer, R.;  et~al.
\newblock Heterogeneous to homogeneous melting transition visualized with
  ultrafast electron diffraction.
\newblock {\em Science} {\bf 2018}, {\em 360},~1451--1455.

\bibitem[Medvedev and Milov(2020)]{medvedev2020nonthermal}
Medvedev, N.; Milov, I.
\newblock Nonthermal phase transitions in metals.
\newblock {\em Scientific reports} {\bf 2020}, {\em 10},~1--9.

\bibitem[Jourdain \em{et~al.}(2021)Jourdain, Lecherbourg, Recoules, Renaudin,
  and Dorchies]{jourdain2021ultrafast}
Jourdain, N.; Lecherbourg, L.; Recoules, V.; Renaudin, P.; Dorchies, F.
\newblock Ultrafast Thermal Melting in Nonequilibrium Warm Dense Copper.
\newblock {\em Physical Review Letters} {\bf 2021}, {\em 126},~065001.

\bibitem[Allen(1987)]{allen1987theory}
Allen, P.B.
\newblock Theory of thermal relaxation of electrons in metals.
\newblock {\em Physical review letters} {\bf 1987}, {\em 59},~1460.

\bibitem[Vorberger and Gericke(2012)]{vorberger2012theory}
Vorberger, J.; Gericke, D.
\newblock Theory of electron-ion energy transfer applied to laser ablation.
\newblock  AIP Conference Proceedings. American Institute of Physics,  2012,
  Vol. 1464, pp. 572--581.

\bibitem[Waldecker \em{et~al.}(2016)Waldecker, Bertoni, Ernstorfer, and
  Vorberger]{waldecker2016electron}
Waldecker, L.; Bertoni, R.; Ernstorfer, R.; Vorberger, J.
\newblock Electron-phonon coupling and energy flow in a simple metal beyond the
  two-temperature approximation.
\newblock {\em Physical Review X} {\bf 2016}, {\em 6},~021003.

\bibitem[Sadasivam \em{et~al.}(2017)Sadasivam, Chan, and
  Darancet]{sadasivam2017theory}
Sadasivam, S.; Chan, M.K.; Darancet, P.
\newblock Theory of thermal relaxation of electrons in semiconductors.
\newblock {\em Physical review letters} {\bf 2017}, {\em 119},~136602.

\bibitem[Maldonado \em{et~al.}(2017)Maldonado, Carva, Flammer, and
  Oppeneer]{maldonado2017theory}
Maldonado, P.; Carva, K.; Flammer, M.; Oppeneer, P.M.
\newblock Theory of out-of-equilibrium ultrafast relaxation dynamics in metals.
\newblock {\em Physical Review B} {\bf 2017}, {\em 96},~174439.

\bibitem[Ono(2018)]{ono2018thermalization}
Ono, S.
\newblock Thermalization in simple metals: Role of electron-phonon and
  phonon-phonon scattering.
\newblock {\em Physical Review B} {\bf 2018}, {\em 97},~054310.

\bibitem[Ritzmann \em{et~al.}(2020)Ritzmann, Oppeneer, and
  Maldonado]{ritzmann2020theory}
Ritzmann, U.; Oppeneer, P.M.; Maldonado, P.
\newblock Theory of out-of-equilibrium electron and phonon dynamics in metals
  after femtosecond laser excitation.
\newblock {\em Physical Review B} {\bf 2020}, {\em 102},~214305.

\bibitem[Miao and Wang(2021)]{miao2021nonequilibrium}
Miao, W.; Wang, M.
\newblock Nonequilibrium effects on the electron-phonon coupling constant in
  metals.
\newblock {\em Physical Review B} {\bf 2021}, {\em 103},~125412.

\bibitem[Zahn \em{et~al.}(2021{\natexlab{a}})Zahn, Jakobs, Windsor, Seiler,
  Vasileiadis, Butcher, Qi, Engel, Atxitia, Vorberger, et~al.]{zahn2021lattice}
Zahn, D.; Jakobs, F.; Windsor, Y.W.; Seiler, H.; Vasileiadis, T.; Butcher,
  T.A.; Qi, Y.; Engel, D.; Atxitia, U.; Vorberger, J.;  et~al.
\newblock Lattice dynamics and ultrafast energy flow between electrons, spins,
  and phonons in a 3d ferromagnet.
\newblock {\em Physical Review Research} {\bf 2021}, {\em 3},~023032.

\bibitem[Zahn \em{et~al.}(2021{\natexlab{b}})Zahn, Jakobs, Seiler, Butcher,
  Engel, Vorberger, Atxitia, Windsor, and Ernstorfer]{zahn2021intrinsic}
Zahn, D.; Jakobs, F.; Seiler, H.; Butcher, T.A.; Engel, D.; Vorberger, J.;
  Atxitia, U.; Windsor, Y.W.; Ernstorfer, R.
\newblock Intrinsic energy flow in laser-excited 3d ferromagnets.
\newblock {\em arXiv preprint arXiv:2110.00525} {\bf 2021}.

\bibitem[Anisimov(1967)]{anisimov1967effect}
Anisimov, S.I.
\newblock Effect of the powerful light fluxes on metals.
\newblock {\em Sov. Phys. Tech. Phys.} {\bf 1967}, {\em 11},~945.

\bibitem[Anisimov \em{et~al.}(1974)Anisimov, Kapeliovich, Perelman,
  et~al.]{anisimov1974electron}
Anisimov, S.; Kapeliovich, B.; Perelman, T.;  et~al.
\newblock Electron emission from metal surfaces exposed to ultrashort laser
  pulses.
\newblock {\em Zh. Eksp. Teor. Fiz} {\bf 1974}, {\em 66},~375--377.

\bibitem[Zahn \em{et~al.}(2021)Zahn, Seiler, Windsor, and
  Ernstorfer]{zahn2021ultrafast}
Zahn, D.; Seiler, H.; Windsor, Y.W.; Ernstorfer, R.
\newblock Ultrafast lattice dynamics and electron--phonon coupling in platinum
  extracted with a global fitting approach for time-resolved polycrystalline
  diffraction data.
\newblock {\em Structural Dynamics} {\bf 2021}, {\em 8}.

\bibitem[Lin \em{et~al.}(2008)Lin, Zhigilei, and Celli]{lin2008electron}
Lin, Z.; Zhigilei, L.V.; Celli, V.
\newblock Electron-phonon coupling and electron heat capacity of metals under
  conditions of strong electron-phonon nonequilibrium.
\newblock {\em Physical Review B} {\bf 2008}, {\em 77},~075133.

\bibitem[Mueller and Rethfeld(2013)]{mueller2013relaxation}
Mueller, B.; Rethfeld, B.
\newblock Relaxation dynamics in laser-excited metals under nonequilibrium
  conditions.
\newblock {\em Physical Review B} {\bf 2013}, {\em 87},~035139.

\bibitem[Petrov \em{et~al.}(2013)Petrov, Inogamov, and
  Migdal]{petrov2013thermal}
Petrov, Y.V.; Inogamov, N.; Migdal, K.P.
\newblock Thermal conductivity and the electron-ion heat transfer coefficient
  in condensed media with a strongly excited electron subsystem.
\newblock {\em JETP letters} {\bf 2013}, {\em 97},~20--27.

\bibitem[Gorbunov \em{et~al.}(2015)Gorbunov, Medvedev, Terekhin, and
  Volkov]{gorbunov2015electron}
Gorbunov, S.; Medvedev, N.; Terekhin, P.; Volkov, A.
\newblock Electron--lattice coupling after high-energy deposition in aluminum.
\newblock {\em Nuclear Instruments and Methods in Physics Research Section B:
  Beam Interactions with Materials and Atoms} {\bf 2015}, {\em 354},~220--225.

\bibitem[Brown \em{et~al.}(2016)Brown, Sundararaman, Narang, Goddard~III, and
  Atwater]{brown2016ab}
Brown, A.M.; Sundararaman, R.; Narang, P.; Goddard~III, W.A.; Atwater, H.A.
\newblock Ab initio phonon coupling and optical response of hot electrons in
  plasmonic metals.
\newblock {\em Physical Review B} {\bf 2016}, {\em 94},~075120.

\bibitem[Medvedev and Milov(2020)]{medvedev2020electron}
Medvedev, N.; Milov, I.
\newblock Electron-phonon coupling in metals at high electronic temperatures.
\newblock {\em Physical Review B} {\bf 2020}, {\em 102},~064302.

\bibitem[Medvedev and Milov(2021)]{medvedev2021contribution}
Medvedev, N.; Milov, I.
\newblock Contribution of inter-and intraband transitions into electron-phonon
  coupling in metals.
\newblock {\em arXiv preprint arXiv:2103.08185} {\bf 2021}.

\bibitem[Ji and Zhang(2016)]{ji2016ab}
Ji, P.; Zhang, Y.
\newblock Ab initio determination of effective electron--phonon coupling factor
  in copper.
\newblock {\em Physics Letters A} {\bf 2016}, {\em 380},~1551--1555.

\bibitem[Migdal \em{et~al.}(2016)Migdal, Petrov, Il‘nitsky, Zhakhovsky,
  Inogamov, Khishchenko, Knyazev, and Levashov]{migdal2016heat}
Migdal, K.; Petrov, Y.V.; Il‘nitsky, D.; Zhakhovsky, V.; Inogamov, N.;
  Khishchenko, K.; Knyazev, D.; Levashov, P.
\newblock Heat conductivity of copper in two-temperature state.
\newblock {\em Applied Physics A} {\bf 2016}, {\em 122},~1--5.

\bibitem[Smirnov(2020)]{smirnov2020copper}
Smirnov, N.
\newblock Copper, gold, and platinum under femtosecond irradiation: Results of
  first-principles calculations.
\newblock {\em Physical Review B} {\bf 2020}, {\em 101},~094103.

\bibitem[Ogitsu \em{et~al.}(2018)Ogitsu, Fernandez-Pa{\~n}ella, Hamel, Correa,
  Prendergast, Pemmaraju, and Ping]{ogitsu2018ab}
Ogitsu, T.; Fernandez-Pa{\~n}ella, A.; Hamel, S.; Correa, A.; Prendergast, D.;
  Pemmaraju, C.; Ping, Y.
\newblock Ab initio modeling of nonequilibrium electron-ion dynamics of iron in
  the warm dense matter regime.
\newblock {\em Physical Review B} {\bf 2018}, {\em 97},~214203.

\bibitem[Caro \em{et~al.}(2015)Caro, Correa, Tamm, Samolyuk, and
  Stocks]{caro2015adequacy}
Caro, A.; Correa, A.; Tamm, A.; Samolyuk, G.D.; Stocks, G.M.
\newblock Adequacy of damped dynamics to represent the electron-phonon
  interaction in solids.
\newblock {\em Physical Review B} {\bf 2015}, {\em 92},~144309.

\bibitem[Migdal \em{et~al.}(2015)Migdal, Il'Nitsky, Petrov, and
  Inogamov]{migdal2015equations}
Migdal, K.; Il'Nitsky, D.; Petrov, Y.V.; Inogamov, N.
\newblock Equations of state, energy transport and two-temperature hydrodynamic
  simulations for femtosecond laser irradiated copper and gold.
\newblock  Journal of Physics: Conference Series. IOP Publishing,  2015, Vol.
  653, p. 012086.

\bibitem[Li and Ji(2022)]{li2022ab}
Li, Y.; Ji, P.
\newblock Ab initio calculation of electron temperature dependent electron heat
  capacity and electron-phonon coupling factor of noble metals.
\newblock {\em Computational Materials Science} {\bf 2022}, {\em 202},~110959.

\bibitem[Wingert \em{et~al.}(2020)Wingert, Singer, Patel, Kukreja, Verstraete,
  Romero, Uhl{\'\i}{\v{r}}, Festersen, Zhu, Glownia, et~al.]{wingert2020direct}
Wingert, J.; Singer, A.; Patel, S.; Kukreja, R.; Verstraete, M.J.; Romero,
  A.H.; Uhl{\'\i}{\v{r}}, V.; Festersen, S.; Zhu, D.; Glownia, J.;  et~al.
\newblock Direct time-domain determination of electron-phonon coupling
  strengths in chromium.
\newblock {\em Physical Review B} {\bf 2020}, {\em 102},~041101.

\bibitem[Milov \em{et~al.}(2018)Milov, Lipp, Medvedev, Makhotkin, Louis, and
  Bijkerk]{milov2018modeling}
Milov, I.; Lipp, V.; Medvedev, N.; Makhotkin, I.A.; Louis, E.; Bijkerk, F.
\newblock Modeling of XUV-induced damage in Ru films: the role of model
  parameters.
\newblock {\em JOSA B} {\bf 2018}, {\em 35},~B43--B53.

\bibitem[Petrov \em{et~al.}(2020)Petrov, Migdal, Inogamov, Khokhlov, Ilnitsky,
  Milov, Medvedev, Lipp, and Zhakhovsky]{petrov2020ruthenium}
Petrov, Y.; Migdal, K.; Inogamov, N.; Khokhlov, V.; Ilnitsky, D.; Milov, I.;
  Medvedev, N.; Lipp, V.; Zhakhovsky, V.
\newblock Ruthenium under ultrafast laser excitation: Model and dataset for
  equation of state, conductivity, and electron-ion coupling.
\newblock {\em Data in brief} {\bf 2020}, {\em 28},~104980.

\bibitem[Hohenberg and Kohn(1964)]{hohenberg1964inhomogeneous}
Hohenberg, P.; Kohn, W.
\newblock Inhomogeneous electron gas.
\newblock {\em Physical review} {\bf 1964}, {\em 136},~B864.

\bibitem[Mermin(1965)]{mermin1965thermal}
Mermin, N.D.
\newblock Thermal properties of the inhomogeneous electron gas.
\newblock {\em Physical Review} {\bf 1965}, {\em 137},~A1441.

\bibitem[Wang \em{et~al.}(1994)Wang, Riffe, Lee, and Downer]{wang1994time}
Wang, X.; Riffe, D.M.; Lee, Y.S.; Downer, M.
\newblock Time-resolved electron-temperature measurement in a highly excited
  gold target using femtosecond thermionic emission.
\newblock {\em Physical Review B} {\bf 1994}, {\em 50},~8016.

\bibitem[B{\'e}villon \em{et~al.}(2014)B{\'e}villon, Colombier, Recoules, and
  Stoian]{bevillon2014free}
B{\'e}villon, E.; Colombier, J.P.; Recoules, V.; Stoian, R.
\newblock Free-electron properties of metals under ultrafast laser-induced
  electron-phonon nonequilibrium: A first-principles study.
\newblock {\em Physical Review B} {\bf 2014}, {\em 89},~115117.

\bibitem[Gonze \em{et~al.}(2009)Gonze, Amadon, Anglade, Beuken, Bottin,
  Boulanger, Bruneval, Caliste, Caracas, C{\^o}t{\'e}, et~al.]{gonze2009abinit}
Gonze, X.; Amadon, B.; Anglade, P.M.; Beuken, J.M.; Bottin, F.; Boulanger, P.;
  Bruneval, F.; Caliste, D.; Caracas, R.; C{\^o}t{\'e}, M.;  et~al.
\newblock ABINIT: First-principles approach to material and nanosystem
  properties.
\newblock {\em Computer Physics Communications} {\bf 2009}, {\em
  180},~2582--2615.

\bibitem[Gonze \em{et~al.}(2016)Gonze, Jollet, Araujo, Adams, Amadon,
  Applencourt, Audouze, Beuken, Bieder, Bokhanchuk, et~al.]{gonze2016recent}
Gonze, X.; Jollet, F.; Araujo, F.A.; Adams, D.; Amadon, B.; Applencourt, T.;
  Audouze, C.; Beuken, J.M.; Bieder, J.; Bokhanchuk, A.;  et~al.
\newblock Recent developments in the ABINIT software package.
\newblock {\em Computer Physics Communications} {\bf 2016}, {\em
  205},~106--131.

\bibitem[Gonze(1997)]{gonze1997first}
Gonze, X.
\newblock First-principles responses of solids to atomic displacements and
  homogeneous electric fields: Implementation of a conjugate-gradient
  algorithm.
\newblock {\em Physical Review B} {\bf 1997}, {\em 55},~10337.

\bibitem[Gonze and Lee(1997)]{gonze1997dynamical}
Gonze, X.; Lee, C.
\newblock Dynamical matrices, Born effective charges, dielectric permittivity
  tensors, and interatomic force constants from density-functional perturbation
  theory.
\newblock {\em Physical Review B} {\bf 1997}, {\em 55},~10355.

\bibitem[Savrasov and Savrasov(1996)]{savrasov1996electron}
Savrasov, S.Y.; Savrasov, D.Y.
\newblock Electron-phonon interactions and related physical properties of
  metals from linear-response theory.
\newblock {\em Physical Review B} {\bf 1996}, {\em 54},~16487.

\bibitem[Perdew \em{et~al.}(1996)Perdew, Burke, and
  Ernzerhof]{perdew1996generalized}
Perdew, J.P.; Burke, K.; Ernzerhof, M.
\newblock Generalized gradient approximation made simple.
\newblock {\em Physical review letters} {\bf 1996}, {\em 77},~3865.

\bibitem[Kohn and Sham(1965)]{kohn1965self}
Kohn, W.; Sham, L.J.
\newblock Self-consistent equations including exchange and correlation effects.
\newblock {\em Physical review} {\bf 1965}, {\em 140},~A1133.

\bibitem[Baroni \em{et~al.}(2001)Baroni, De~Gironcoli, Dal~Corso, and
  Giannozzi]{baroni2001phonons}
Baroni, S.; De~Gironcoli, S.; Dal~Corso, A.; Giannozzi, P.
\newblock Phonons and related crystal properties from density-functional
  perturbation theory.
\newblock {\em Reviews of modern Physics} {\bf 2001}, {\em 73},~515.

\bibitem[Allen and Dynes(1975)]{Allen1975}
Allen, P.B.; Dynes, R.C.
\newblock Transition temperature of strong-coupled superconductors reanalyzed.
\newblock {\em Phys. Rev. B} {\bf 1975}, {\em 12},~905--922.
\newblock
  doi:{\changeurlcolor{black}\href{https://doi.org/10.1103/physrevb.12.905}{\detokenize{10.1103/physrevb.12.905}}}.

\bibitem[Molina and White(2021)]{molina2021molecular}
Molina, J.M.; White, T.G.
\newblock A Molecular Dynamics Study of Laser-Excited Gold.
\newblock {\em arXiv preprint arXiv:2101.00499} {\bf 2021}.

\bibitem[Ivanov and Zhigilei(2003)]{ivanov2003combined}
Ivanov, D.S.; Zhigilei, L.V.
\newblock Combined atomistic-continuum modeling of short-pulse laser melting
  and disintegration of metal films.
\newblock {\em Physical Review B} {\bf 2003}, {\em 68},~064114.

\bibitem[Tamm \em{et~al.}(2018)Tamm, Caro, Caro, Samolyuk, Klintenberg, and
  Correa]{tamm2018langevin}
Tamm, A.; Caro, M.; Caro, A.; Samolyuk, G.; Klintenberg, M.; Correa, A.A.
\newblock Langevin dynamics with spatial correlations as a model for
  electron-phonon coupling.
\newblock {\em Physical review letters} {\bf 2018}, {\em 120},~185501.

\end{thebibliography}
\end{document}